\documentclass[aps,showpacs,preprintnumbers,amsmath, amssymb]{revtex4}

\usepackage{float}
\usepackage{graphics,epsfig}
\usepackage{graphicx}
\usepackage{dcolumn}
\usepackage{bm}

\begin{document}
\baselineskip=0.8 cm
\title{{\bf Estimation precision of the acceleration for a two-level atom coupled to fluctuating vacuum electromagnetic fields}}

\author{Mengge Zheng$^{1}$, Baoyuan Yang$^{1}$, Zixu Zhao$^{1}$\footnote{Corresponding author. zhao$_{-}$zixu@yeah.net}}
\affiliation{$^{1}$School of Science, Xi'an University of Posts and Telecommunications, Xi'an 710121, China}

\vspace*{0.2cm}
\begin{abstract}
\baselineskip=0.6 cm
\begin{center}
{\bf Abstract}
\end{center}

In open quantum systems, we study the quantum Fisher information of acceleration for a uniformly accelerated two-level atom coupled to fluctuating electromagnetic fields in the Minkowski vacuum. With the time evolution, for the initial atom state parameter $\theta\neq\pi$, the quantum Fisher information can exist a maximum value and a local minimum value before reaching a stable value. In addition, in a short time, the quantum Fisher information varies with the initial state parameter, and the quantum Fisher information can take a maximum value at $\theta=0$. The quantum Fisher information may exist two peak values at a certain moment. These features are different from the massless scalar fields case. With the time evolution, $F_{max}$ firstly increases, then decreases, and finally, reaches the same value. However, $F_{max}$ will arrive at a stable maximum value for the case of the massless scalar fields. Although the atom response to the vacuum fluctuation electromagnetic fields is different from the case of massless scalar fields, the quantum Fisher information eventually reaches a stable value.

\end{abstract}

\maketitle
\newpage
\vspace*{0.2cm}

\section{Introduction}

Every quantum system is inevitably subjected to one environment because it at least interacts with vacuum fluctuations. There has been accumulated interest in studying vacuum fluctuations since the Lamb shift \cite{Lamb,Bethe} and the Casimir (and Casimir-Polder ) effects \cite{Casimir1,Casimir2} were discovered. The Unruh effect showed that, a uniformly accelerated observer perceives the Minkowski vacuum as a thermal bath of Rindler particles \cite{Rindler,Fulling,Hawking1,Hawking2,Davies1,DeWitt,Unruh}. This is the result of the quantum field theory \cite{BirrellDavies}. The modification of vacuum fluctuations is induced by the acceleration of a two-level atom.
In Ref. \cite{Letaw}, the authors considered the second quantization of the free scalar field in rotating coordinates and calculated the spectrum of vacuum fluctuations for an orbiting observer using these coordinates. They found that the spectrum of vacuum fluctuations is composed of the usual zero-point energy plus a contribution arising from the observer's acceleration. In Ref. \cite{Bell}, the possibility of using accelerated electrons to exhibit the quantum field theoretic relation between acceleration and temperature has been considered. The authors have examined further the connection between the Unruh effect and the radiative excitations of electrons in the storage rings \cite{Bell2}. Audretsch and M\"{u}ller considered a uniformly accelerated atom coupled to a massless scalar quantum field \cite{AM1,AM2}, and their results are consistent with the Unruh effect. The authors in Ref. \cite{AMM} have studied the generalized Unruh effect and Lamb shift for an atom in a circular motion. Passante has studied the radiative level shifts of an accelerated hydrogen atom and the Unruh effect in quantum electrodynamics, and it was shown that the effect of electromagnetic vacuum fluctuations on atomic level shifts is not totally equivalent to that of a thermal field \cite{Passante}. The direct detection of the Unruh effect (or Unruh-like effect) is difficult. Therefore, one can study the effect in the frame of the estimation theory.

Parameter metrology is an active topic in quantum information theory since experimental errors and uncertainties are unavoidable. The estimation theory presents the method to obtain the fundamental precision bounds of parameter estimation and find the optimal measurement strategies \cite{Helstrom,Holevo}. Based on the best measuring strategy \cite{Holevo}, Bu\v{z}ek, Derka, and Massar studied the optimal quantum clock \cite{Buzek}, and the experimental realization of such an optimal quantum clock using trapped ions has been discussed. Giovannetti, Lloyd, and Maccone studied Quantum-Enhanced measurements \cite{Giovannetti1}, and pointed out a general framework that encompasses most cases in which quantum effects enable an increase in precision when estimating a parameter \cite{Giovannetti2}. They also studied the effects of noise and experimental imperfections \cite{Giovannetti3}. Quantum Fisher information (QFI) places a fundamental limit to the accuracy of quantum estimation; therefore, it plays an extremely important role in quantum metrology. Li and Luo proposed an entanglement criterion in terms of quantum Fisher information, and illustrated the significance of this criterion by showing that it can reveal entanglement undetectable by the variance method \cite{Li}. There has attracted considerable attention recently \cite{Zhong,Sun,Wang,Jin,Tian,Rajabpour,Gessner,Zhao,Zhao2}. In order to compare to the result of the quantum Fisher information for an accelerated two-level atom coupled to a massless scalar field in the Minkowski vacuum \cite{Zhao}, we would like to know the case of the vacuum electromagnetic fields. In this work, we will study the QFI of acceleration for a uniformly accelerated two-level atom coupled with fluctuating vacuum electromagnetic fields.

The structure of this paper is as follows. In Sec. II, we review the open quantum system and quantum Fisher information. In Sec. III, for a uniformly accelerated two-level atom coupled to electromagnetic fields in the Minkowski vacuum, we analyze the estimation precision of acceleration by calculating the QFI of acceleration in free space. We will summarize our results in the last section.

\section{Open quantum system and quantum Fisher information}

We consider a two-level atom coupled to a bath of fluctuating quantum electromagnetic fields in vacuum. The Hamiltonian of the whole system takes the form
\begin{equation}
H=H_A+H_F+H_I.
\end{equation}
Here $H_A={1\over 2}\,\hbar\omega_0\sigma_3,$ is the Hamiltonian of the atom, where
$\sigma_3$ is the Pauli matrix, and $\omega_0$ is the energy-level spacing of the atom. $H_F$ is the Hamiltonian of the electromagnetic field, of which the explicit form is not relevant here. In the multipolar coupling scheme~\cite{CPP}, the interaction Hamiltonian $H_I$ takes the
form $H_I(\tau)=-e\,\textbf{r} \cdot
\textbf{E}(x(\tau))=-e\sum_{mn}\textbf{r}_{mn}\cdot
\textbf{E}(x(\tau))\sigma_{mn},$ where {\it e} is the electron
electric charge, $e\,\bf r$ the atomic electric dipole moment, and ${\bf E}(x)$ the electric field strength.

The initial state of the whole system is characterized by the total
density matrix $\rho_{tot}=\rho(0) \otimes |0\rangle\langle0|$, in
which $\rho(0)$ is the initial reduced density matrix of the atom,
and $|0\rangle$ is the vacuum state of the field. In the frame of
the atom, the evolution in the proper time $\tau$ of the total
density matrix $\rho_{tot}$ satisfies
\begin{equation}
\frac{\partial\rho_{tot}(\tau)}{\partial\tau}=-{\frac{i}{\hbar}}[H,\rho_{tot}(\tau)]\;.
\end{equation}
We assume that the interaction between the field and the atom is
weak. The evolution of the reduced density matrix $\rho(\tau)$ in the limit of weak coupling can be written in the
Kossakowski-Lindblad form~\cite{Gorini,Lindblad,Benatti1}
\begin{equation}
{\partial\rho(\tau)\over \partial \tau}= -{\frac{i}{\hbar}}\big[H_{\rm eff},\,
\rho(\tau)\big]
 + {\cal L}[\rho(\tau)]\ ,
\end{equation}
where
\begin{equation}
{\cal L}[\rho]={1\over2} \sum_{i,j=1}^3
a_{ij}\big[2\,\sigma_j\rho\,\sigma_i-\sigma_i\sigma_j\, \rho
-\rho\,\sigma_i\sigma_j\big]\ .
\end{equation}
The matrix $a_{ij}$ and the effective Hamiltonian $H_{\rm eff}$ are determined by the Fourier and Hilbert transforms of the correlation functions,
\begin{equation}
G^{+}(x-y)={\frac{e^2}{\hbar^2}} \sum_{i,j=1}^3\langle +|r_i|-\rangle\langle -|r_j|+\rangle\,\langle0|E_i(x)E_j(y)|0 \rangle\;,
\end{equation}
which are defined as follows:
\begin{equation}
{\cal G}(\lambda)=\int_{-\infty}^{\infty} d\tau \,
e^{i{\lambda}\tau}\, G^+\big(x(\tau)\big)\; ,
\end{equation}
\begin{equation}
{\cal K}(\lambda)= \frac{P}{\pi i}\int_{-\infty}^{\infty} d\omega\
\frac{ {\cal G}(\omega) }{\omega-\lambda} \;.
\end{equation}
Therefore, the coefficients of the Kossakowski matrix $a_{ij}$ can be expressed as
\begin{equation}
a_{ij}=A\delta_{ij}-iB
\epsilon_{ijk}\delta_{k3}-A\delta_{i3}\delta_{j3}\;,
\end{equation}
in which
\begin{equation}\label{ab}
A=\frac{1}{4}[{\cal G}(\omega_0)+{\cal G}(-\omega_0)]\;,\;~~~
B=\frac{1}{4}[{\cal G}(\omega_0)-{\cal G}(-\omega_0)]\;.
\end{equation}
The effective Hamiltonian $H_{\rm eff}$ contains a correction term,
the so-called Lamb shift, and one can show that it is given
by replacing $\omega_0$  in $H_s$ with a renormalized energy-level
spacing $\Omega$ as follows:
\begin{equation}\label{heff}
H_{\rm eff}=\frac{1}{2}\hbar\Omega\sigma_3={\hbar\over 2}\left\{\omega_0+\frac{i}{2}[{\cal
K}(-\omega_0)-{\cal K}(\omega_0)]\right\}\sigma_3\;.
\end{equation}

Making use of the ansatz that the initial state of the atom $|\psi(0)\rangle=\cos\frac{\theta}{2}|+\rangle+e^{i\phi}\sin\frac{\theta}{2}|-\rangle$, we obtain the evolution of Bloch vector
\begin{align}\label{omega}
&\omega_1(\tau)=\sin\theta \cos(\Omega\tau+\phi)e^{-2 A\tau}\;,\nonumber\\
&\omega_2(\tau)=\sin\theta \sin(\Omega\tau+\phi)e^{-2 A\tau}\;,\nonumber\\
&\omega_3(\tau)=\cos\theta e^{-4 A\tau}-\frac{B}{A}\left(1-e^{-4 A\tau}\right)\;.
\end{align}

It is well known that, in the quantum metrology, the QFI provides a lower bound to the mean-square error in the estimation by Cram\'{e}r-Rao inequality~\cite{Helstrom,Holevo,Fisher,Cramer,Bures,Uhlmann,Hubner,Braunstein}
\begin{equation}\label{1}
\rm{Var}(X)\geq\frac{1}{N F_X}\;,
\end{equation}
with the number of repeated measurements $N$. Here $F_X$ represents the QFI of parameter $X$, which can be calculated in terms of the symmetric logarithmic derivative operator by
\begin{equation}
F_X=\textrm{Tr}\,(\rho(X)L^2)\;,
\end{equation}
with the symmetric logarithmic derivative Hermitian operator $L$ which satisfies the equation $\partial_X \rho(X)=[\rho(X)L+L\rho(X)]/2$. For a two-level system, the state can be expressed in the Bloch sphere as $\rho=(I+\bm{\omega}\cdot\bm{\sigma})/2$, here $\bm{\omega}=(\omega_1,\omega_2,\omega_3)$ is the Bloch vector and $\bm{\sigma}=(\sigma_1,\sigma_2,\sigma_3)$ denotes the Pauli matrices. Thus, the QFI of parameter $X$ can be written as \cite{Zhong}
\begin{equation}\label{FX}
   F_X=\left\{
    \begin{array}{l}
    |\partial_X\bm{\omega}|^2+\frac{(\bm{\omega}\cdot\partial_X\bm{\omega})^2}{1-|\bm{\omega}|^2}\;,\;\;\,|\bm{\omega}|<1\;,  \\
    |\partial_X\bm{\omega}|^2\;,\;\;\;\;\;\;\;\;\;\;\;\;\;\;\;\;\;\;|\bm{\omega}|=1\;. \\
    \end{array}
    \right.
\end{equation}

\section{Quantum estimation of acceleration}

We considered a uniformly accelerated two-level atom coupled to a electromagnetic field in the Minkowski vacuum. The trajectory of the atom can be described as
\begin{eqnarray}
t{(\tau)}={\frac{c}{a}}\sinh \frac{a\tau}{c}\;,~~~
x{(\tau)}={\frac{c^2}{a}}\cosh \frac{a\tau}{c}\;,~~~
y{(\tau)}=0\;,~~~
z{(\tau)}=0\label{trajl}\;.
\end{eqnarray}

We need the field correlation function, which can be calculated by using the two-point function of the electric field
\begin{eqnarray}
\langle E_i(x(\tau))E_j(x(\tau'))\rangle&=&{\frac{\hbar c}{4\pi^2\varepsilon_0}}(\partial
_0\partial_0^\prime\delta_{ij}-\partial_i\partial_j^\prime)
{\frac{1}{(x-x')^2+(y-y')^2+(z-z')^2-(ct-ct'-i\varepsilon)^2}}\;.
\end{eqnarray}

Using the trajectory (\ref{trajl}), one can obtain the field  correlation function
\begin{equation}
G^{+}(x,x')={\frac{e^2|\langle -|\mathbf{r}|+\rangle|^2}{16\pi^2\varepsilon_0\hbar c^7}}\frac{a^4}{\sinh^4[\frac{a}{2c}(\tau-\tau'-i\varepsilon)]} .
\end{equation}

Therefore, the Fourier transform of the field correlation function is
\begin{eqnarray}
{\cal G}(\lambda)={\frac{e^2|\langle -|\mathbf{r}|+\rangle|^2\lambda^3}{6\pi\varepsilon_0\hbar c^3}}(1+\frac{a^2}{c^2\lambda^2})(1+\coth \frac{\pi c \lambda}{a})\;.
\end{eqnarray}

Therefore, we obtain
\begin{eqnarray}
A={\frac{\gamma_0}{4}}(1+\frac{a^2}{c^2\omega_0^2})\frac{e^{2\pi c\omega_0 /a}+1}{e^{2\pi c\omega_0 /a}-1}\;,B={\frac{\gamma_0}{4}}(1+\frac{a^2}{c^2\omega_0^2}),
\end{eqnarray}
where $\gamma_0={e^2|\langle -|\mathbf{r}|+\rangle|^2\omega_0^3}/(3\pi\varepsilon_0\hbar c^3)$ is the spontaneous emission rate. Therefore, we can calculate the QFI. It should be noted that the expression is too long to exhibit here. In the following discussion, we use $\tau\rightarrow \tilde{\tau}\equiv{\tau}{\gamma_0}$, $a\rightarrow \tilde{a}\equiv{a}/{(c\omega_0)}$.  For simplicity, $\tilde{\tau}$ and $\tilde{a}$ will be written as $\tau$ and $a$.

\begin{figure}[ht]
\includegraphics[scale=0.45]{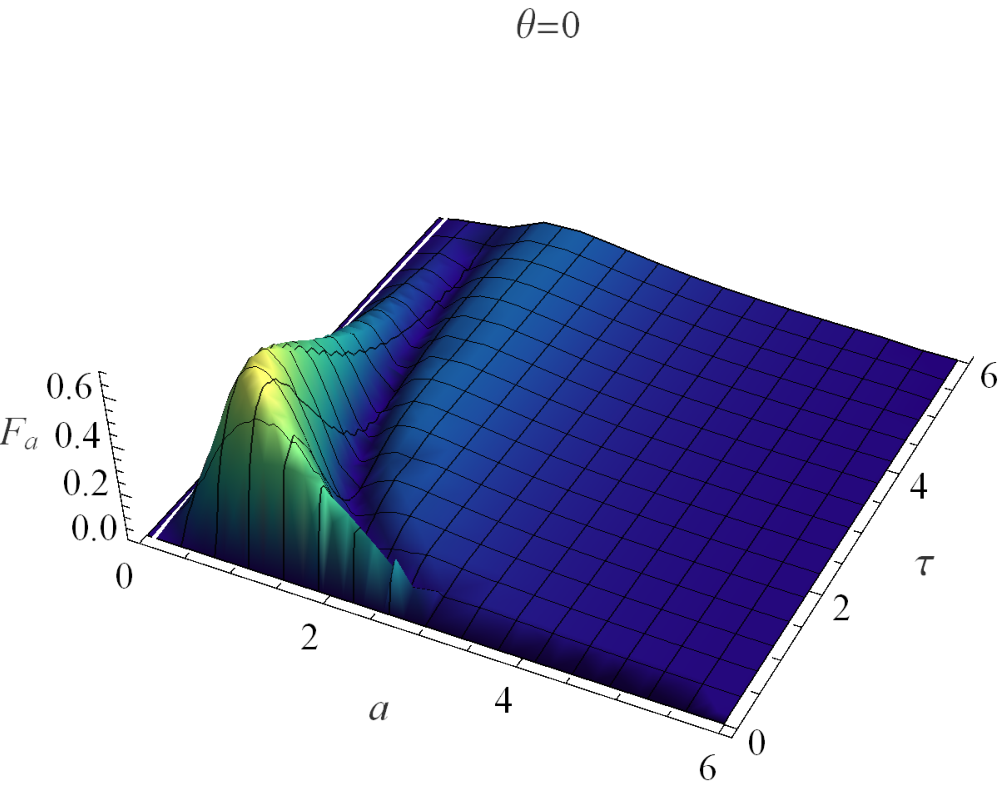}\vspace{0.0cm}
\includegraphics[scale=0.45]{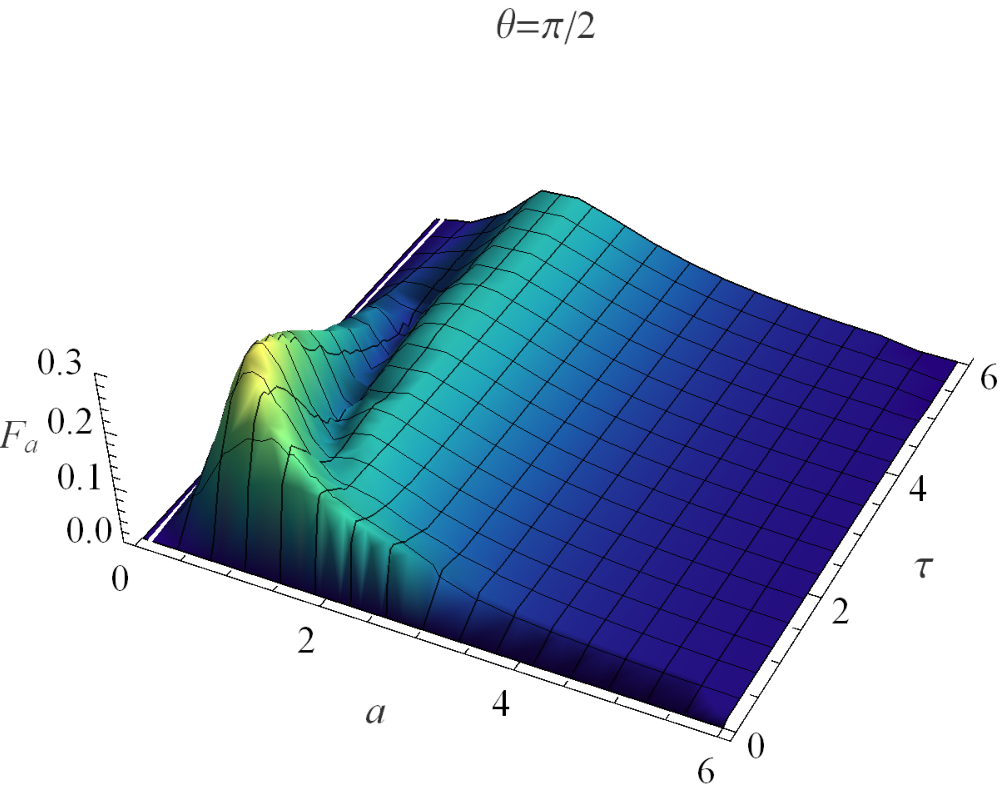}\vspace{0.0cm}
\includegraphics[scale=0.45]{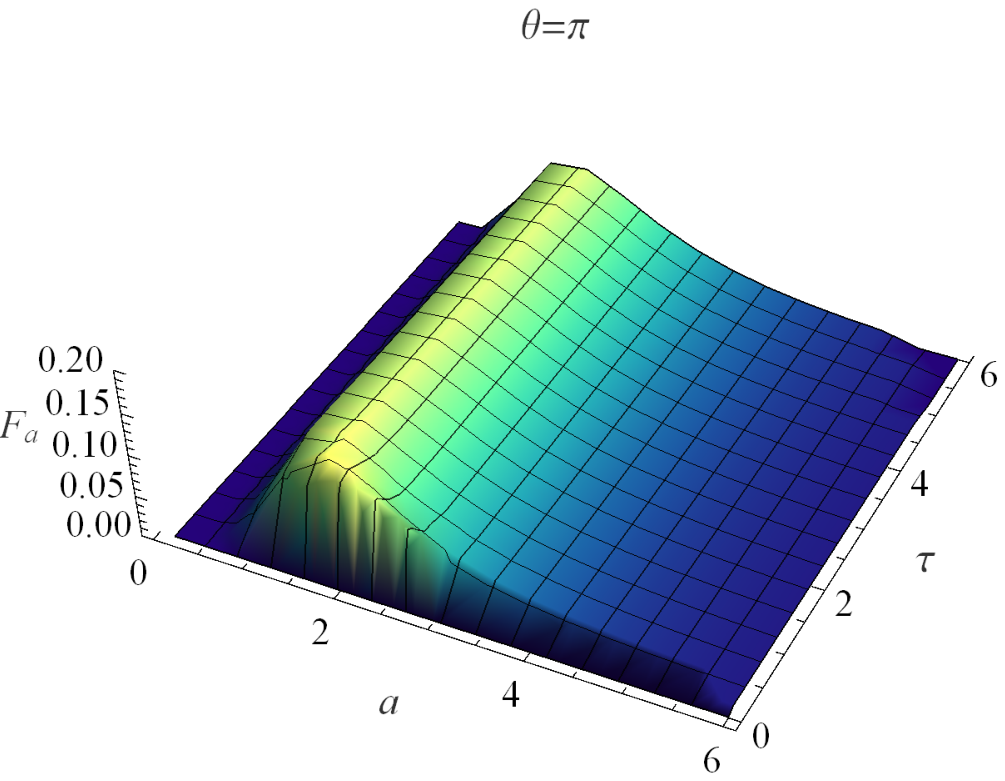}\vspace{0.0cm}
\caption{\label{3DFatau} The QFI of acceleration as a function of the evolution time $\tau$ and acceleration $a$ with the initial atom state $\theta=0$, $\theta=\pi/2$ and $\theta=\pi$.}
\end{figure}

\begin{figure}[ht]
\includegraphics[scale=0.45]{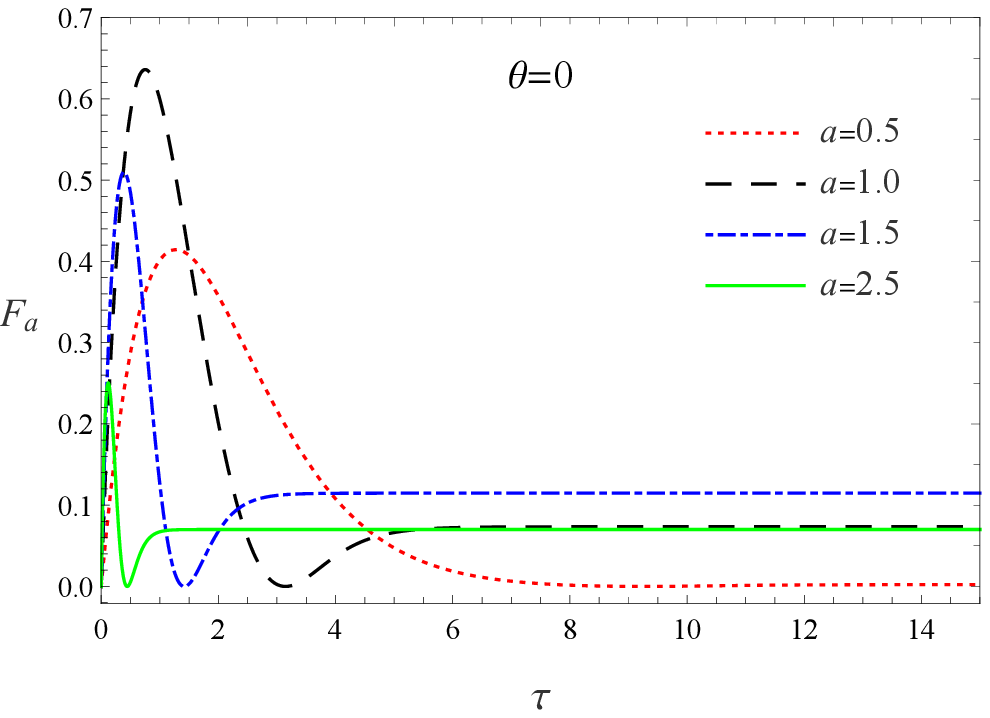}\vspace{0.0cm}
\includegraphics[scale=0.45]{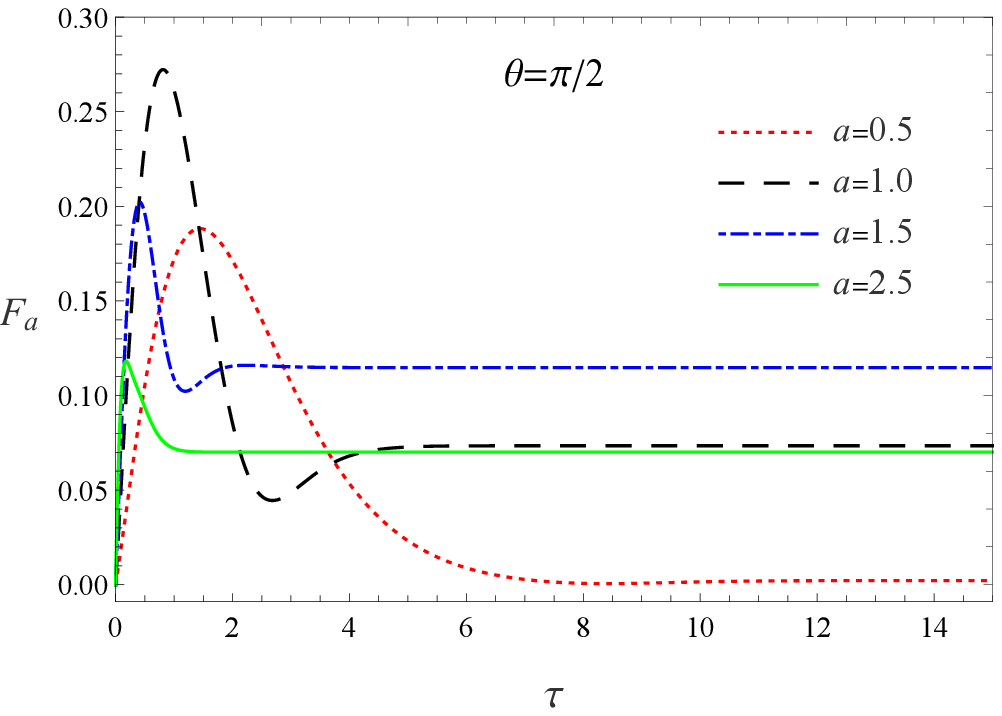}\vspace{0.0cm}
\includegraphics[scale=0.45]{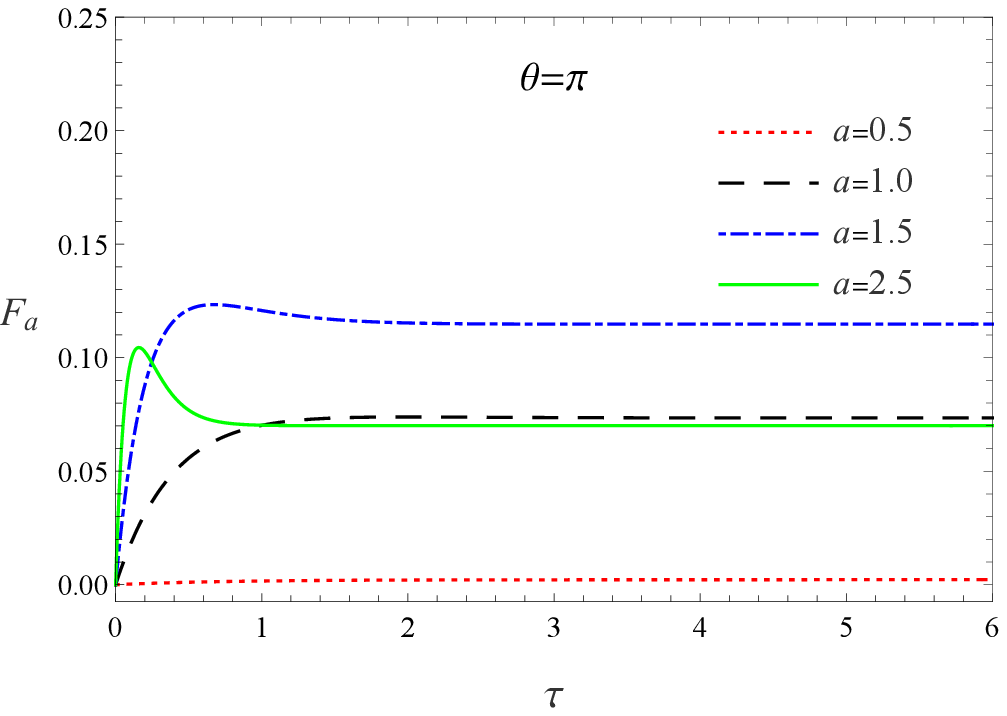}\vspace{0.0cm}
\caption{\label{Ftau} The QFI of acceleration as a function of the evolution time $\tau$ for different acceleration $a$ with the initial atom state $\theta=0$, $\theta=\pi/2$ and $\theta=\pi$. We take $a=0.5$ (red dotted line), $a=1.0$ (black dashed line), $a=1.5$ (blue dot-dashed line) and $a=2.5$ (green solid line).}
\end{figure}

In Fig. \ref{3DFatau}, we describe the QFI of acceleration as a function of the evolution time $\tau$ and the acceleration $a$ with $\theta=0$, $\theta=\pi/2$ and $\theta=\pi$. The behavior that the initial atom state is in the ground state ($\theta=\pi$) is different from the case of other initial states. We can also observe the feature from Fig. \ref{Ftau}, where we depict the QFI of acceleration as a function of $\tau$ for different $a$ with $\theta=0$, $\theta=\pi/2$ and $\theta=\pi$. For $\theta=0$ and $\theta=\pi/2$, we find that the QFI can exist a maximum value and a local minimum value before reaching a stable value. For a smaller $a$, the QFI firstly increases, then decreases, and finally, goes to zero. This is different from the result in the massless scalar fields case \cite{Zhao}. There is an optimal detection time at which the highest precision for estimating the acceleration can be achieved. For different values of $a$, the peak values in general are different. When the initial atom state is in the ground state ($\theta=\pi$), the situation is quite different. For a larger $a$, the QFI increases at the beginning, then decreases, and finally keeps it steady for a long enough time. For a smaller $a$, the QFI increase firstly, and finally reach a stable value. The result is similar to the massless scalar fields case.

\begin{figure}[ht]
\includegraphics[scale=0.45]{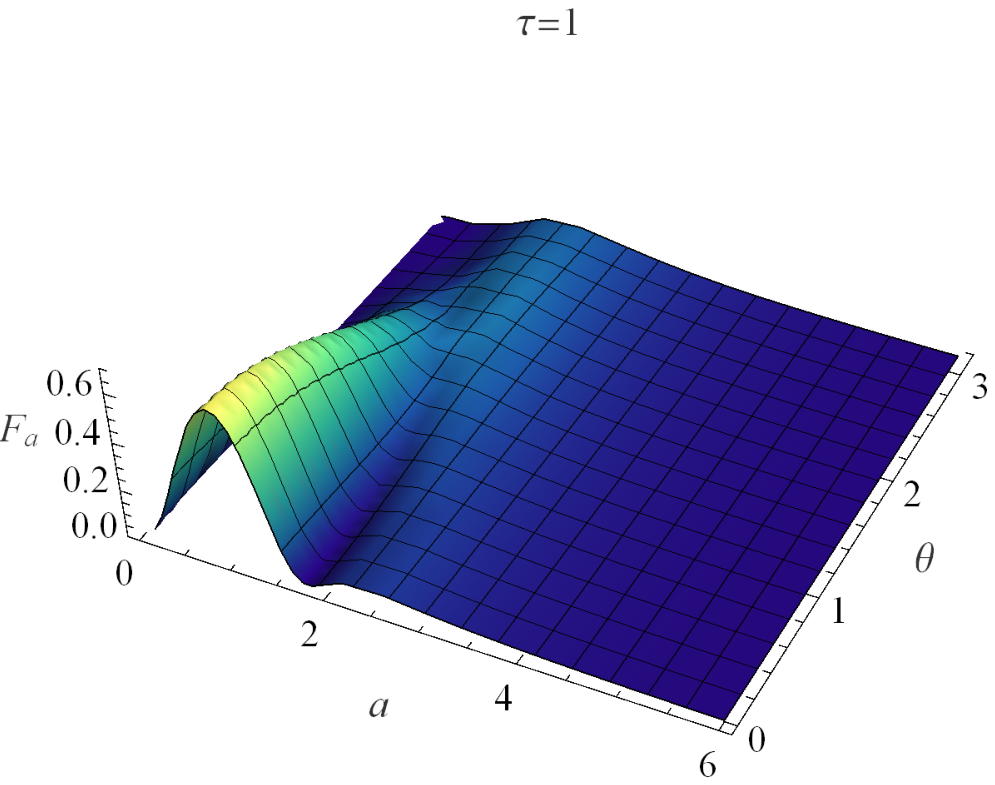}\vspace{0.0cm}
\includegraphics[scale=0.45]{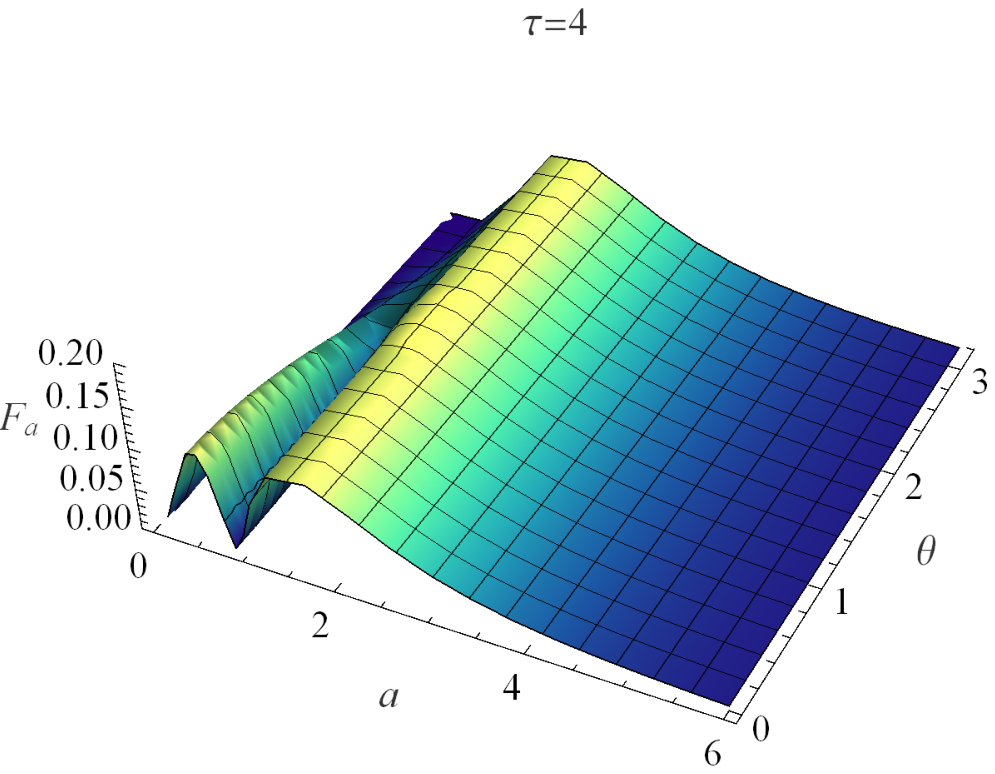}\vspace{0.0cm}
\includegraphics[scale=0.45]{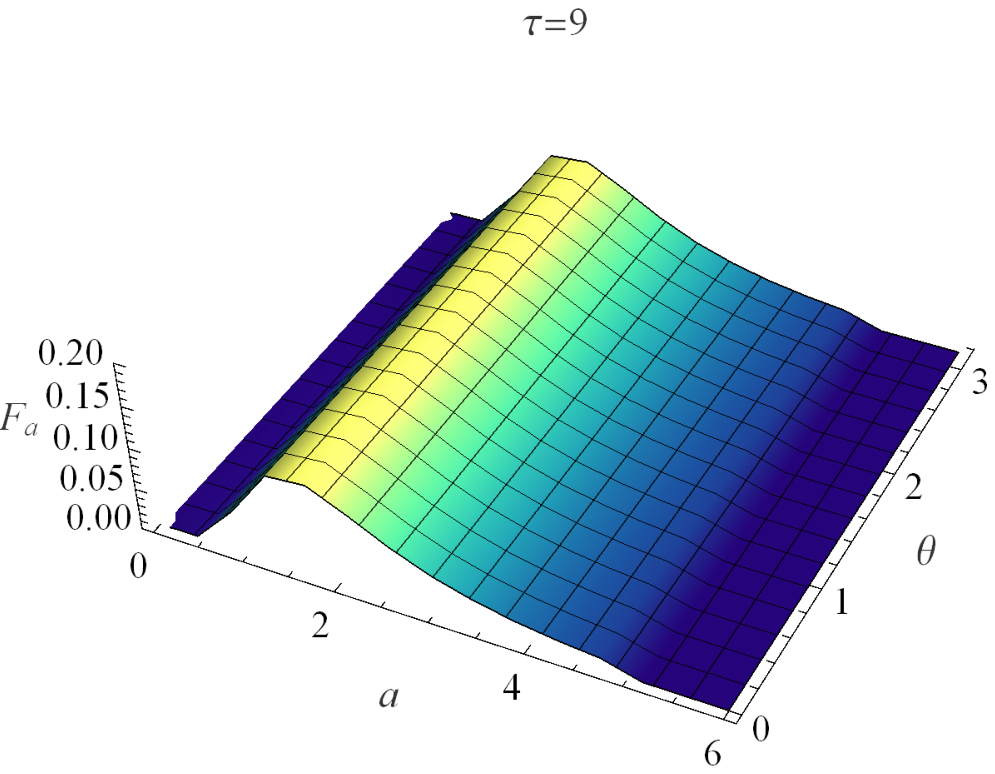}\vspace{0.0cm}
\caption{\label{3DFatheta} The QFI of acceleration as a function of $a$ and $\theta$ with $\tau=1$, $\tau=4$ and $\tau=9$.}
\end{figure}

\begin{figure}[ht]
\includegraphics[scale=0.45]{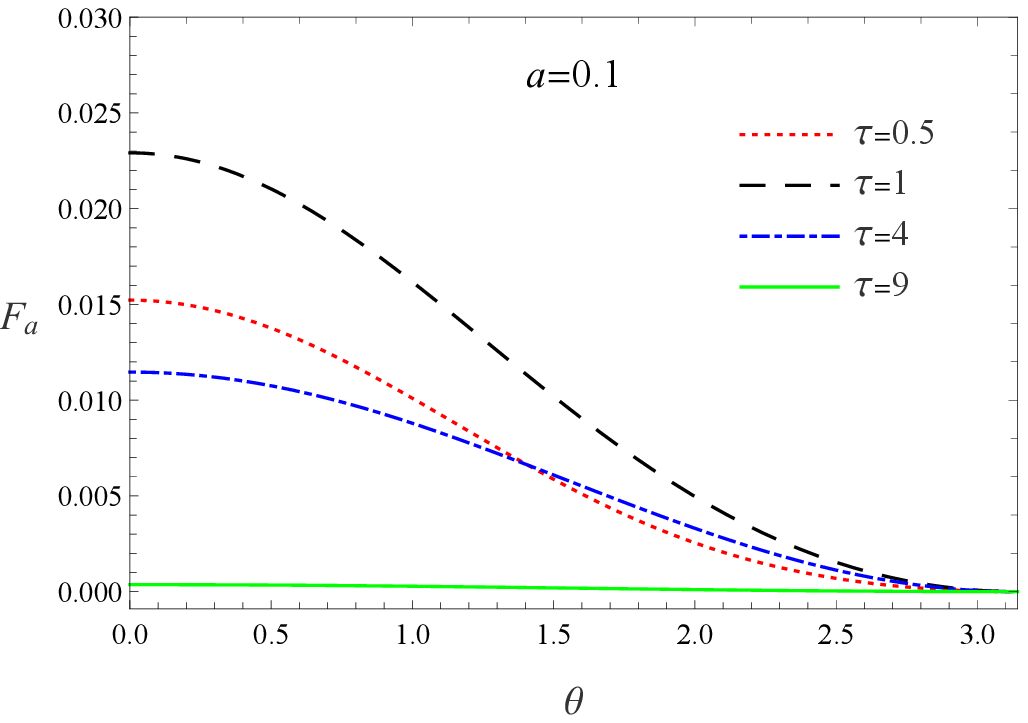}\vspace{0.0cm}
\includegraphics[scale=0.45]{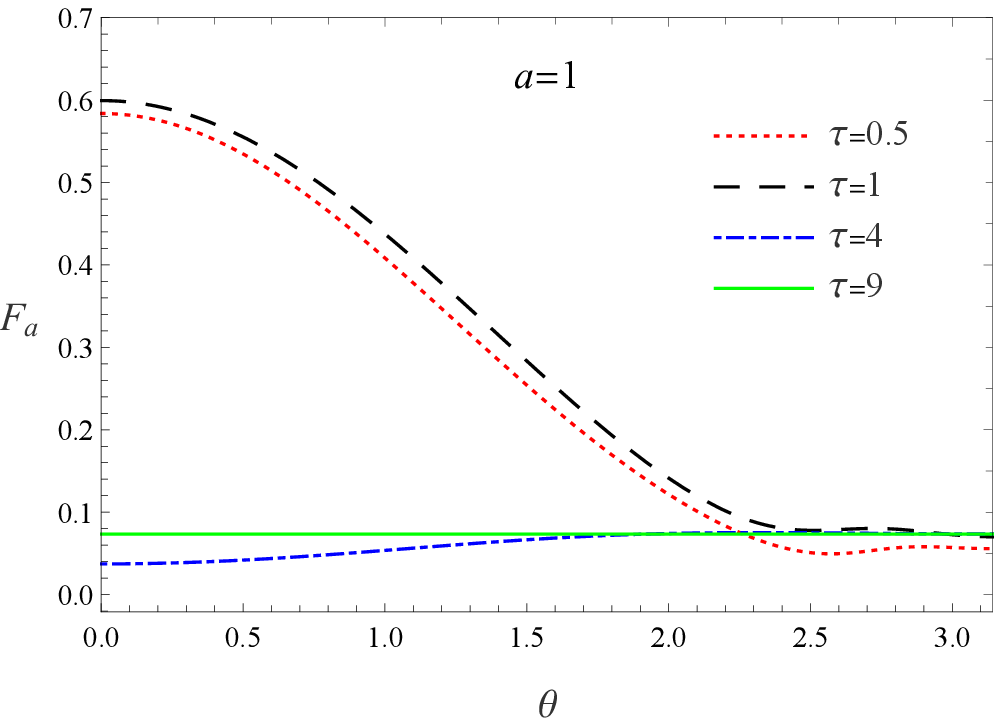}\vspace{0.0cm}
\includegraphics[scale=0.45]{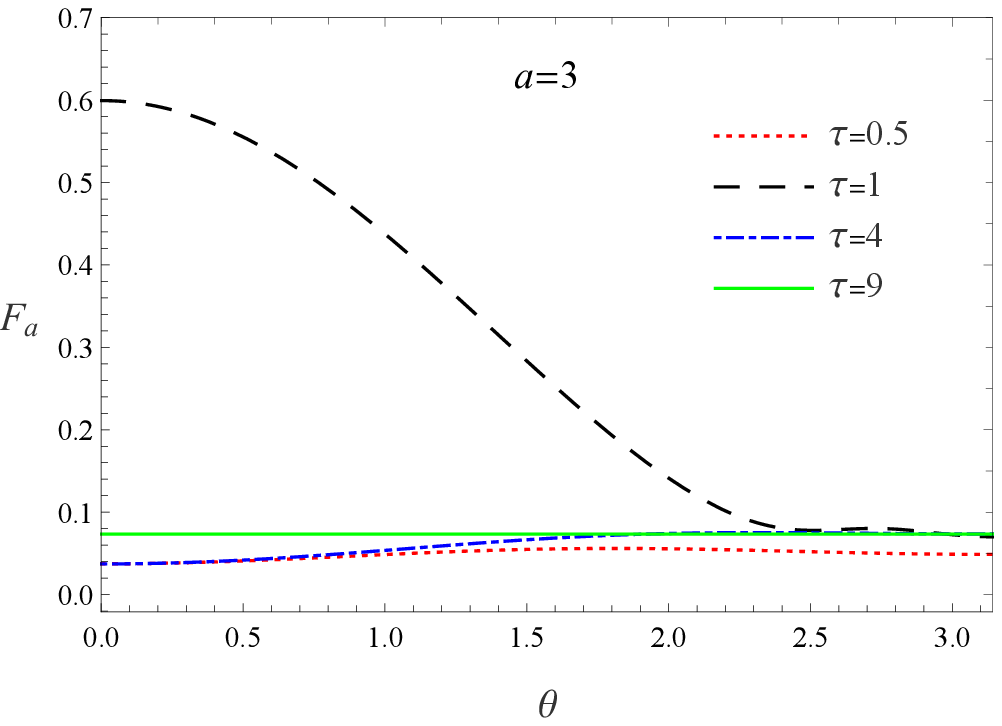}\vspace{0.0cm}
\caption{\label{Ftheta} The QFI of acceleration as a function of $\theta$ for different $\tau$ with $a=0.1$, $a=1$ and $a=3$. We take $\tau=0.5$ (red dotted line), $\tau=1$ (black dashed line), $\tau=4$ (blue dot-dashed line) and $\tau=9$ (green solid line).}
\end{figure}

In Fig. \ref{3DFatheta}, we plot the QFI of acceleration as a function of the acceleration $a$ and the initial state parameter $\theta$. We find that, in a short time, the QFI varies with the initial state parameter, and the QFI can take maximum value when the initial atom state is in the excited state ($\theta=0$). The atom initial states response to the vacuum fluctuation electromagnetic fields is different from the case of massless scalar fields \cite{Zhao}. The maximum sensitivity in the estimation for the acceleration can be obtained by the initial preparation of the atom in the excited state. The QFI goes to the stable value beyond a certain time. However, the stable value in general is not the maximum value. We can also observe the feature from Fig. \ref{Ftheta}. In Fig. \ref{Ftheta}, we depict the QFI of acceleration as a function of the initial atom state $\theta$ for different $\tau$ with $a=0.1$, $a=1$ and $a=3$. One can obtain a high estimation precision for the acceleration by selecting the excited state at a very short time.

\begin{figure}[ht]
\includegraphics[scale=0.45]{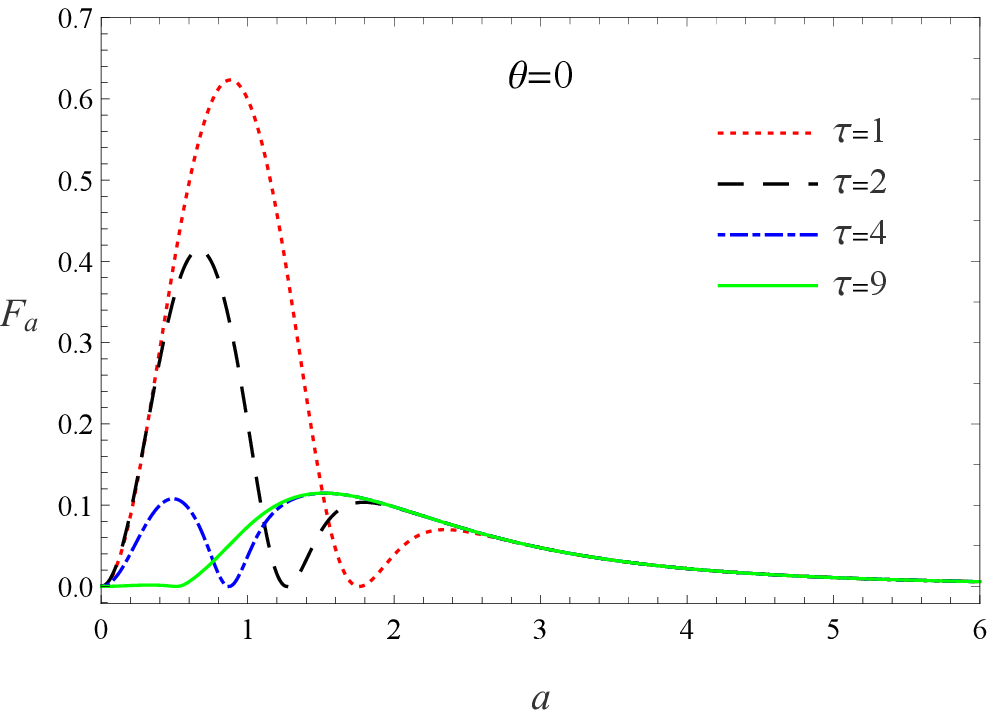}\vspace{0.0cm}
\includegraphics[scale=0.45]{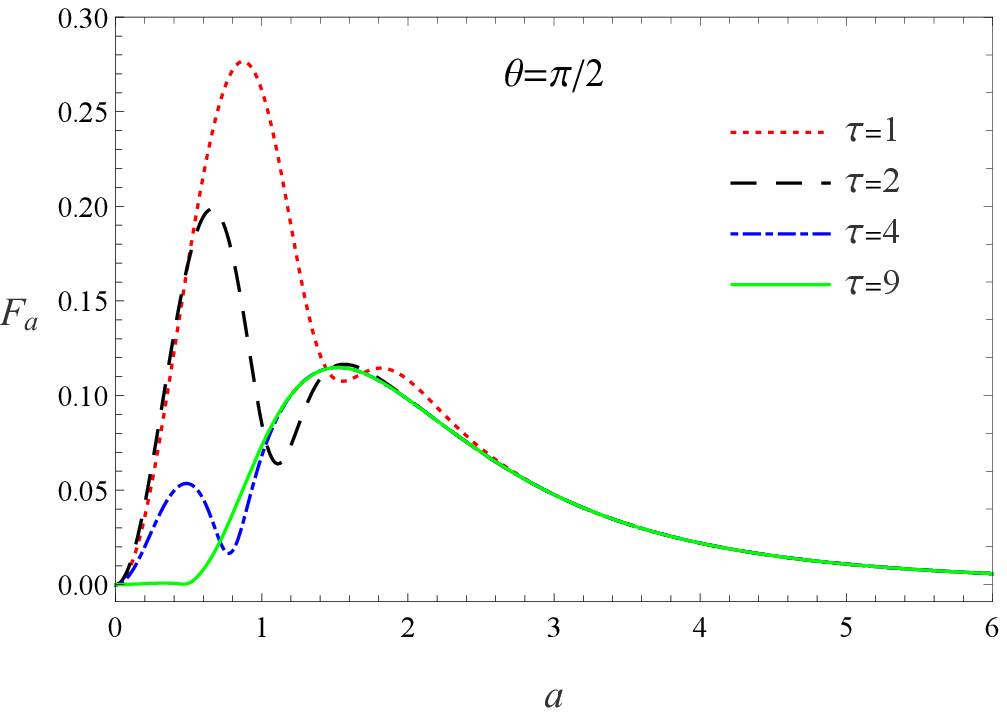}\vspace{0.0cm}
\includegraphics[scale=0.45]{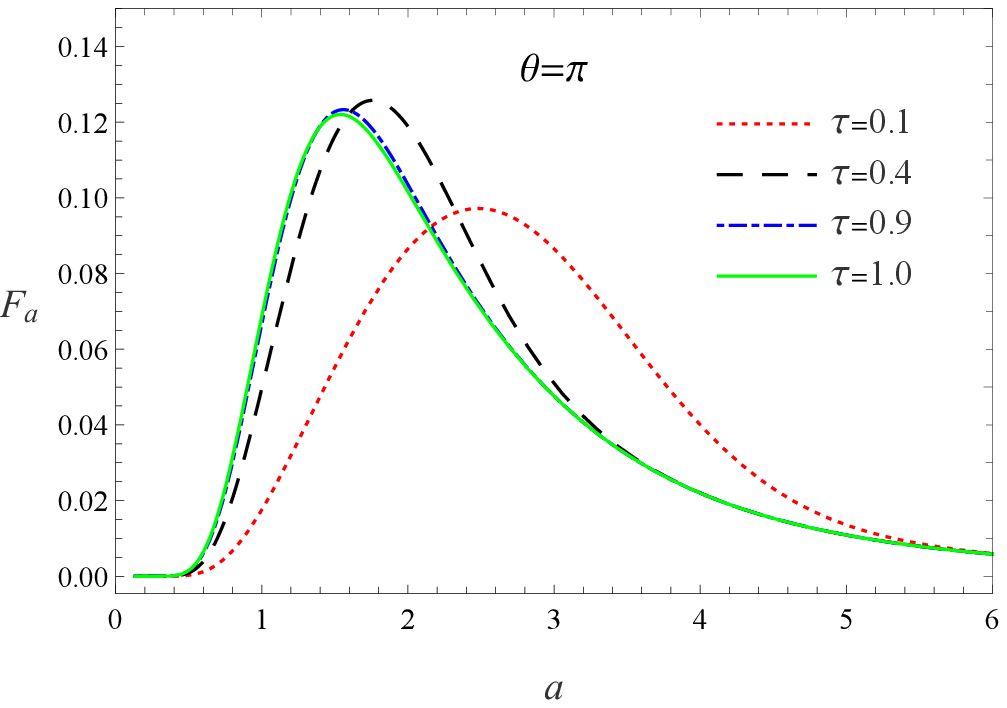}\vspace{0.0cm}
\caption{\label{Fa} The QFI of acceleration as a function of $a$ for different $\tau$ with $\theta=0$, $\theta=\pi/2$ and $\theta=\pi$.}
\end{figure}

In Fig. \ref{Fa}, we plot the QFI of acceleration as a function of $a$ for different $\tau$ with $\theta=0$, $\theta=\pi/2$ and $\theta=\pi$. For $\theta=0$ and $\theta=\pi/2$, we observe that the QFI may exist two peak values at a certain moment. For the case of two peak values, the higher peak value first appears in a smaller $a$, and then appears in a larger $a$. With the evolution of time, the first peak decreases and disappears, and the second peak reaches a stable value. However, there only exists one peak value for $\theta=\pi$. The peak is shifted to the left with the time evolution and finally tends to be stable.

\begin{figure}[ht]
\includegraphics[scale=0.47]{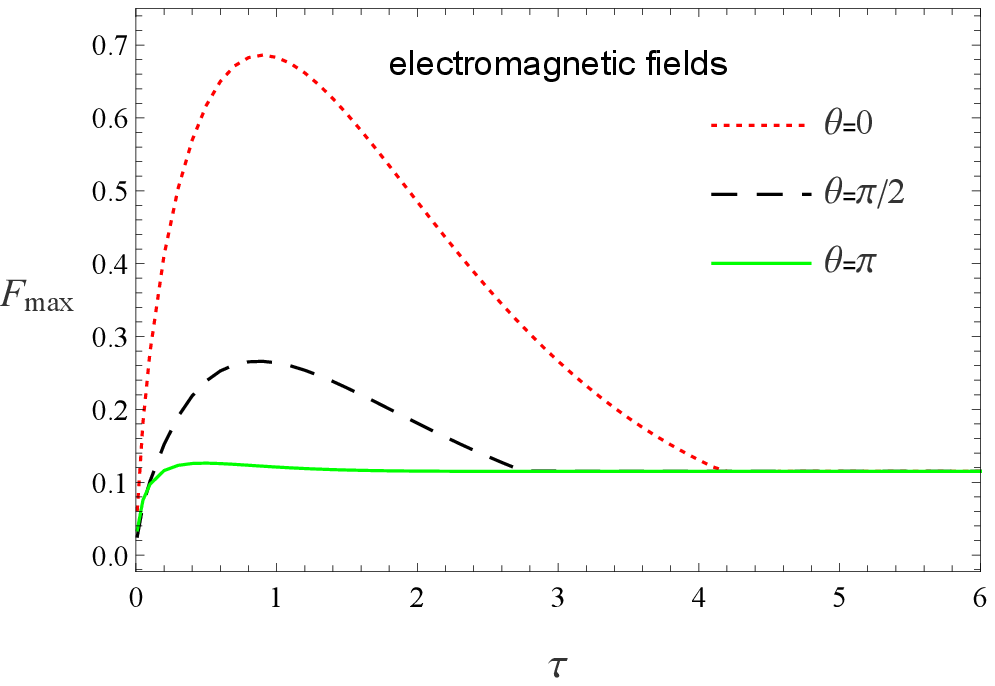}\vspace{0.0cm}
\includegraphics[scale=0.47]{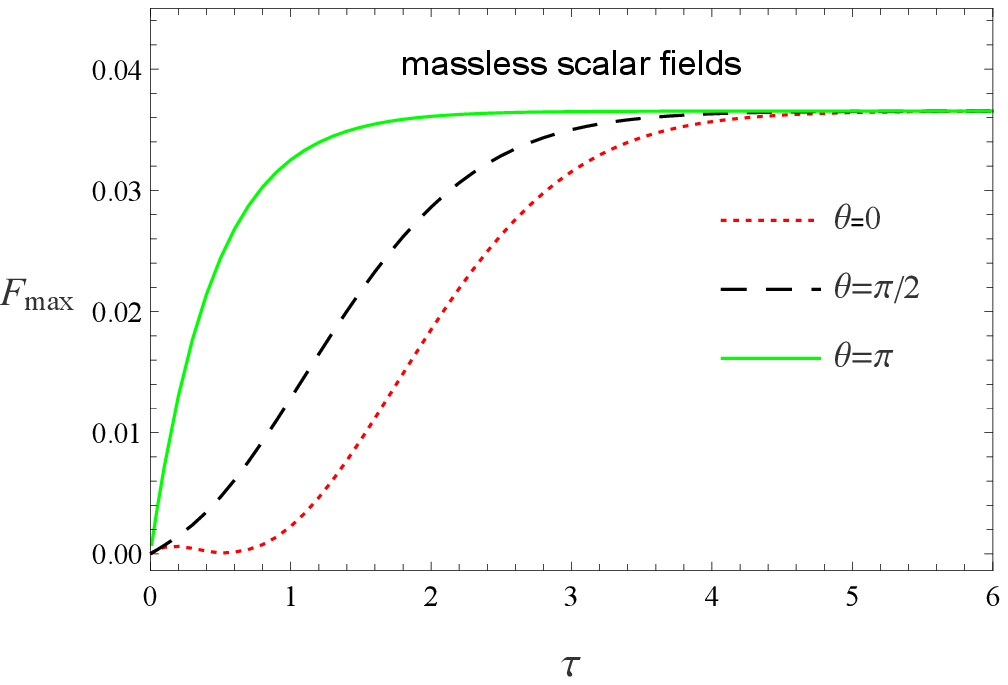}\vspace{0.0cm}
\caption{\label{tausFmax} $F_{max}$ as a function of $\tau$ with different $\theta$. We take $\theta=0$ (red dotted line), $\theta=\pi/2$ (black dashed line) and $\theta=\pi$ (green solid line).}
\end{figure}

From the above discussion, we observe that the situation of the electromagnetic fields is quite different from the massless scalar fields. Now, we describe $F_{max}$ as a function of $\tau$ for the electromagnetic fields in the left panel of Fig. \ref{tausFmax}. $F_{max}$ denotes the maximum value of $F_a$ in Fig. \ref{Fa}. With the increase of $\tau$, $F_{max}$ firstly increases, then decreases, and finally, reaches the same value. In order to obtain a larger QFI, we need to select a small $\tau$. We also depict $F_{max}$ as a function of $\tau$ for the case of the massless scalar fields in the right panel of Fig. \ref{tausFmax}. We observe that $F_{max}$ will arrive at a stable maximum value. Different from the electromagnetic fields, the maximum value can be obtained for a longer time. The QFI is independent of the initial state of the atom for a long time. The feature is similar for the massless scalar fields case and electromagnetic fields.

\section{conclusion}

In the open quantum systems, we have studied the QFI of acceleration for a uniformly accelerated two-level atom coupled to fluctuating vacuum electromagnetic fields. For the atom initial state parameter $\theta\neq \pi$, i.e., preparing the atom not in the ground state, we found that the QFI may exist a maximal value and a local minimum value before reaching a stable value, which is different from the result in the massless scalar fields case. For a larger $a$ with $\theta=\pi$, the QFI firstly increases, then decreases, and finally, keeps it steady for a long enough time with the time evolution. For a smaller $a$, the QFI firstly increase, and finally reach a stable value. The behavior that the initial atom state is in the ground state is different from the case of other initial states. In a short time, we found that the QFI varies with the initial state parameter, and the QFI can take a maximum value at $\theta=0$. The QFI will reach stable value beyond a certain time. However, the stable value is not the maximum. From the pictures of $F_{a}-a$, we found that the QFI may exist two peak values at a certain moment. The higher peak value appears in a smaller $a$ at the beginning, and then appears in a larger $a$. With the time evolution, the first peak decreases and disappears, and the second peak reaches a stable value. However, there exists no more than one peak value for $\theta=\pi$. The peak is shifted to the left with the time evolution and finally tends to be stable. The atom initial states response to the vacuum fluctuation electromagnetic fields is different from the case of massless scalar fields. The QFI is independent of the initial state of the atom for a long time, which is similar to the massless scalar fields case. For the fluctuating vacuum electromagnetic fields, with the time evolution, $F_{max}$ firstly increases, then decreases, and finally, reaches the same value. However, we observed that $F_{max}$ will arrive at a stable maximum value for the case of the massless scalar fields. There exists a great difference between the electromagnetic fields case and the massless scalar fields case in a short time, but the quantum Fisher information eventually reaches a stable value.

\begin{acknowledgments}

This work was supported by the National Natural Science Foundation of China under Grant No. 11705144.

\end{acknowledgments}

\end{document}